\documentclass[12pt]{iopart}
\usepackage{graphicx} 

\begin{document}

\bibliographystyle{nar}
\newcommand{\NVO}{$\rm Ni_3V_2O_8$}
\newcommand{\CVO}{$\rm Co_3V_2O_8$}
\newcommand{\MVO}{$\rm M_3V_2O_8$}
\newcommand{\Kag}{Kagom\'{e}}
\newcommand{\afm}{antiferromagnetic}
\newcommand{\fm}{ferromagnetic}

\title{Magnetic phase diagrams of the \Kag\ staircase compounds \CVO\ and \NVO}

\author{N~R~Wilson, O~A~Petrenko, G~Balakrishnan}

\address{University of Warwick, Department of Physics, Coventry, CV4 7AL, United Kingdom}

\ead{nicola.r.wilson@warwick.ac.uk}

\begin{abstract}
An extensive low temperature magnetisation study of high quality single crystals of the \Kag\ staircase compounds \NVO\ and \CVO\ has been performed, and the $H-T$ phase diagrams have been determined from these measurements. The magnetisation and susceptibility curves for \CVO\ are analysed in terms of their compatibility with the different ferromagnetic and antiferromagnetic structures proposed for this compound. For \NVO, the phase diagram is extended to magnetic fields higher than previously reported; for a field applied along the $a$-axis, the low temperature incommensurate phase is found to close at around 90~kOe.
\end{abstract}


\section{Introduction}
The magnetic effects caused by the \Kag\ type lattice have interested theorists and experimentalists for many years. The two dimensional layers of corner sharing triangles possess macroscopic degeneracy when each vertex is occupied by a Heisenberg spin interacting \afm ally with its nearest neighbours. This causes a high degree of magnetic frustration and the system remains a disordered spin liquid down to zero temperature. 

Samples that possess a perfect form of this structure are difficult to find, and so experimentalists are turning to structures with variations of the \Kag\ lattice. The \Kag\ staircase systems \MVO~\cite{Rogado1,RogadoCu} with M=Ni, Co or Cu have a structure of magnetic atoms that is based on the \Kag\ lattice. Their structure is different in that the layers are not two dimensional but are buckled in and out of the plane. This variation means that the symmetry of the system is reduced and further neighbour interactions are much more significant. These changes partially relieve the frustration and allow long-range order to be established.

The magnetic properties of \NVO\ have been characterised using magnetisation, specific heat and neutron diffraction techniques~\cite{Lawes, Kenzelmann}. The system passes from a paramagnetic phase to a high temperature incommensurate phase at 9.1~K. The next transition is at 6.3~K where the system enters a low temperature incommensurate phase. At 3.9~K the system then enters a commensurate phase with remnant structures, then finally a purely commensurate \afm\ phase below 2~K. An additional interest in \NVO\ is associated with the low-temperature ferroelectricity observed in this compound \cite{Lawesferro} and the related complex behaviour caused by the magnetoelectric interactions \cite{Kenzelmann}.

 It has been suggested that the Co system passes through similar phase transitions~\cite{Szymczak}, but our recent neutron powder diffraction study has shown otherwise~\cite{WilsonGEM}. At 11~K, \CVO\ moves from a paramagnetic state to an incommensurate \afm\ state. At 6~K there is a transition to a ferromagnetic state, in which the system remains down to the lowest temperature studied of 1.7~K. The Curie-Weiss temperature for this system is approximately 15~K. Here, we present the temperature and field dependence of the magnetisation measured in the various magnetic phases of both compounds for different directions of applied magnetic field. For \CVO, the magnetic phase diagram is presented here for the first time and for \NVO, a new phase boundary is shown.
 
\section{Experimental details}
The \Kag\ staircase compounds \NVO\ and \CVO, where Ni has S=1 and Co has S=3/2, have the space group Cmca with lattice parameters \cite{Sauerbrei} $a$=5.936(4)~\AA, $b$=11.420(6)~\AA\ and $c$=8.240(5)~\AA\ for \NVO\ and $a$=6.030(4)~\AA, $b$=11.486(2)~\AA\ and $c$=8.312(5)~\AA\ for \CVO. The atoms form layers of edge-sharing $\rm MO_6$ octahedra, separated by non-magnetic $\rm VO_4$ tetrahedra. The samples were prepared using the floating zone technique previously described~\cite{Balakrishnan}. Single crystals of both samples of approximately 2~cm{$^3$} in volume were produced with mosaic spreads of less than 1{$^\circ$}. An Oxford Instruments Vibrating Sample Magnetometer (VSM) and a Quantum Design SQUID magnetometer were used to measure the temperature and field dependence of the magnetisation in these compounds. The samples were investigated with fields applied along each of the three main crystallographic directions, aligned with an accuracy of at least 3{$^\circ$}. The phase transitions were tracked in the temperature interval between 1.4 and 12~K in a magnetic field range of up to {$\pm$}120~kOe. 

\section{Results and discussion}

\begin{figure}
\begin{center}
\includegraphics[width=1\columnwidth]{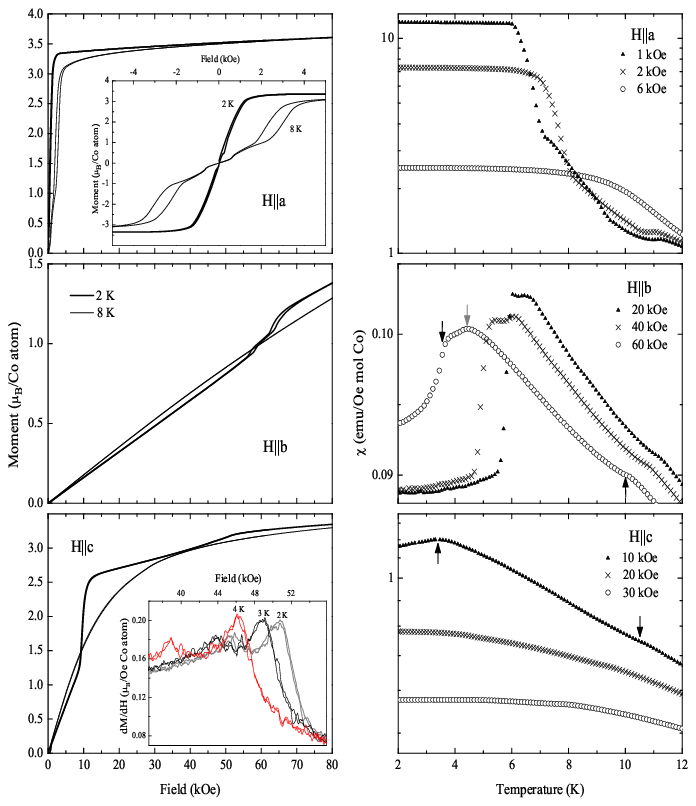}
\caption{\label{fig:Co_hyst}
Left panel: Characteristic magnetisation curves for each of the three main crystallographic directions of \CVO. The 2~K data lie within the \fm\ phase, and the 8~K data in the \afm\ phase. For $H \parallel a$, the inset of the top panel shows there to be no significant hysteresis around zero field in the ferromagnetic phase (within the 50~Oe field resolution of our measurements) and a considerable hysteresis observed around 2.7~kOe in the \afm\ phase. For $H \parallel c$, the derivatives of magnetisation, $dM/dH$, indicate more definitely the location of phase transition fields, as can be seen in the inset on the bottom panel. Right panel: Characteristic susceptibility curves for each of the three main crystallographic directions performed with various applied fields. The black arrows indicate the locations of the  paramagnetic to \afm\ or \afm\ to \fm\ phase transitions. The grey arrow points to the feature in the $\chi(T)$ curve associated with the locking of the incommensurate \afm\ structure into a particular value of the magnetic propagation vector (see main text for details).}
\end{center}
\end{figure}

A range of characteristic magnetisation and susceptibility curves measured in \CVO\ is shown in Figure~\ref{fig:Co_hyst}. The $M(H)$ curve for a field applied along the easy $a$-axis at 2~K shows a very rapid increase caused by a small field -- the magnetic moment is practically saturated at 2~kOe. From the value of magnetisation at $H=80$~kOe, the lower limit for the average Co moment was found to be 3.6~{$\mu_B$}. From the inset of this figure, it is evident that there is no significant remanent magnetisation in zero field (less than 4\% of the saturation magnetisation) and no significant hysteresis, within the measurement field resolution of 50~Oe. From the neutron diffraction data \cite{WilsonGEM} it has been seen that in the ferromagnetic state, the system is not completely ordered; the magnetic moment on one of the two Co sites only has a value of approximately 1.81(4)~{$\mu_B$}. In order to fully understand this magnetisation process, one should take into consideration not just the trivial movement of domain walls but also the possible elongation of existing magnetic moments in an applied magnetic field.

In the \afm\ phase at 8~K, it is again very easy to saturate the magnetisation. In the small field range before saturation, the system clearly passes through a region of hysteresis centred around 2.7~kOe indicating the presence of an additional phase transition. The phase transitions are also clearly seen as anomalies in the $\chi(T)$ curves at low fields. However, by 6~kOe, the susceptibility curve is very smooth, making it impossible to pin-point the transition temperature. Specific heat measurements may prove more useful for plotting the phase diagram for $H\parallel a$, as has been demonstrated for the \NVO\ phase diagram~\cite{Lawes}. 

\begin{figure}
\begin{center}
\includegraphics[width=0.6\columnwidth]{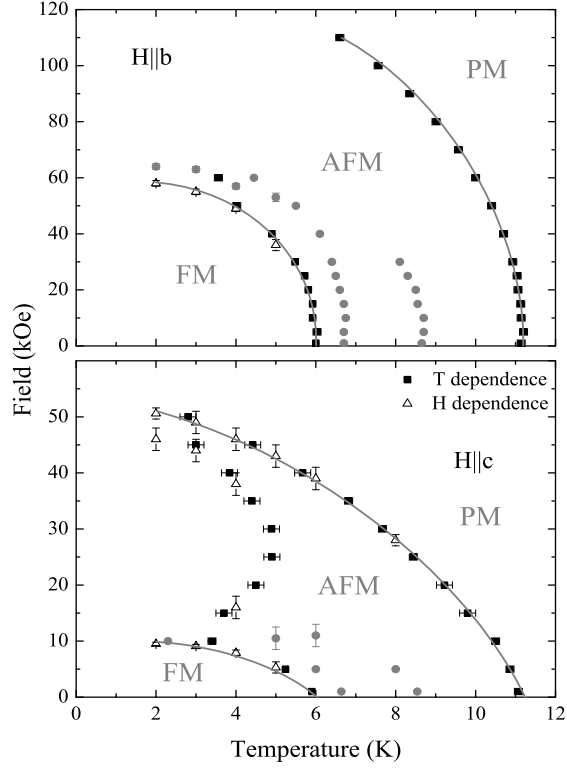}
\caption{\label{fig:Co_phase} Magnetic phase diagram of \CVO\ for magnetic field applied along two main crystallographic directions. Magnetisation temperature and field curves (closed and open symbols) were used to follow the transitions, lines are plotted as guides to the eye. The grey symbols represent anomalies in the magnetisation curves that correspond to the locking and unlocking of the propagation vector within the incommensurate \afm\ phase. Note the different scales of field for each direction.}
\end{center}
\end{figure}

From the magnetisation curves shown for $H\parallel b$, it is obvious that the (010) direction is the hard axis of the system. In a field of 80~kOe, the average magnetic moment is less than 40\% of that found for a field applied along the $a$-axis. The $\chi(T)$ curves measured for these two directions differ by a factor of almost 100~\cite{Balakrishnan, correction}. In the ferromagnetic phase at 2~K, the $M(H)$ curve shows two step-like hysteretic anomalies at 58 and 64~kOe. These features, and others are easily followed across $\chi(T)$ curves measured in different applied fields. From comparison with neutron powder diffraction data it is obvious that the anomaly observed at around 11~K corresponds to the transition from a paramagnetic state to an \afm\ state. The neutron data also shows that the feature at around 6~K corresponds to the transition from an \afm\ state to a ferromagnetic state. These two transitions are shown by black arrows on the $\chi(T)$ curve measured at 60~kOe. The grey arrow corresponds to an anomaly associated with the unlocking of the propagation vector from (0,0.5,0) within the \afm\ phase.

In the ferromagnetic phase at 2~K, the saturation magnetic moment approaches that for $H\parallel a$. The $M(H)$ curve for $H\parallel c$ shows multiple features at 9.5, 46 and 51~kOe. To observe all of these, the derivative of the magentisation must be considered. These irregularities can also be followed in the $\chi(T)$ curves.

All features observed in the $\chi(T)$ and $M(H)$ curves for $H\parallel b$ and $H\parallel c$ are summarised in the magnetic phase diagram in Figure~\ref{fig:Co_phase}. The reduced symmetry and large number of competing interactions in the system contribute to the complex nature of this diagram. It comprises a high temperature paramagnetic phase, an intermediate \afm\ phase and a low temperature ferromagnetic phase. These phases evolve differently as the temperature is reduced and the field increased along different directions. There is added complexity from the anomalies associated with the locking and unlocking of the magnetic propagation vector. 

\begin{figure}
\begin{center}
\includegraphics[width=0.75\columnwidth]{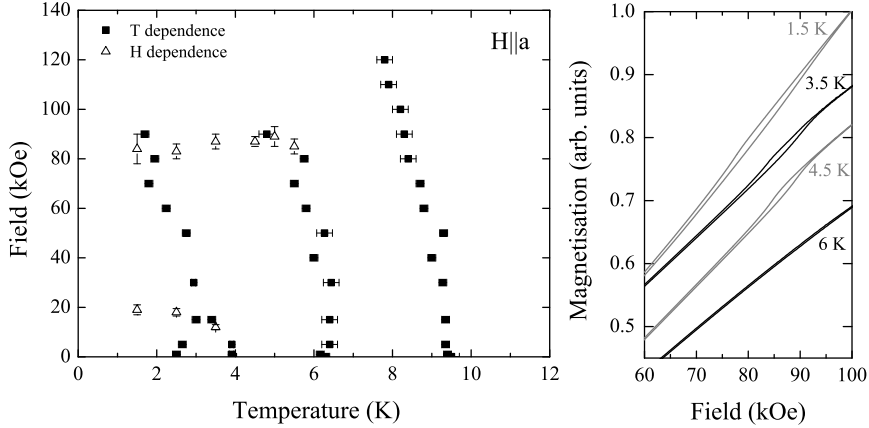}
\caption{\label{fig:Ni_phase} Left panel: Magnetic phase diagram of \NVO\ for magnetic field applied along the $a$-axis. Magnetisation temperature curves (solid symbols) and field curves (open symbols) were used to follow the transitions. Right panel: Field dependence of the magnetisation measured within the low temperature incommensurate phase. The hysteresis apparent at the lower temperatures is no longer observed by 6 K. The curves are offset along the y axis for clarity.}
\end{center}
\end{figure}

The ordering temperature decreases significantly with applied magnetic field, for $H\parallel c$ for example, $T_N(40~$kOe$)\approx 1/2\;T_N(1~$kOe$)$. Such a dramatic decrease in the ordering temperature is unusual even for highly frustrated systems.

A similar range of measurements has been performed on \NVO, confirming the phase diagram reported by Lawes {\it et al.}~\cite{Lawes}. The diagram indicates that five phases exist at low temperatures: paramagnetic, high temperature incommensurate, low temperature incommensurate, and two commensurate phases. Along the $b$ and $c$-axes, the low temperature incommensurate phase is entirely enclosed by 80 and 30~kOe repsectively, but open along the $a$-axis. Here, Figure~\ref{fig:Ni_phase} (left panel) shows the phase diagram for an applied field parallel to the $a$-axis with an extension in magnetic field, up to 120~kOe. The $M(H)$ curves collected between 1.5 and 6~K show slight hysteresis at approximately 90~kOe (Figure~\ref{fig:Ni_phase} right panel). This is evidence for the closing of the low temperature incommensurate phase not previously reported. In addition to the closing of this phase, the phase diagram here also contains a further phase boundary at around 20~kOe at the lowest temperature. This tranistion is evident from the large hysteresis observed in the $M(H)$ curves measured at 1.5 and 2~K.

\section{Conclusions}
The magnetisation measurements performed on the \Kag\ staircase compound \CVO\ show there to be no significant net magnetisation in zero field or hysteresis in the low temperature ferromagnetic phase. The transitions from a paramagnetic state to an \afm\ phase, and then to a ferromagnetic phase are followed in temperature and applied field. The magnetic phase diagram, shown here for the first time, is complicated further by anomalies caused by the locking and unlocking of the propagation vector within the \afm\ phase. The magnetic phase diagram of the closely related compound \NVO\ has been extended in magnetic field to show a new boundary that closes the low temperature incommensurate phase when the magnetic field is applied along the $a$-axis.

We are grateful to L~C~Chapon and B~F{\aa}k for discussion, and to the EPSRC for financial support. 

\section*{References}

\bibliography{Wilson_561_HFM2006_bib}

\begin{thebibliography}{10}
\bibitem{correction}
Reference 9 uses a different notation to this paper and to reference 6 for
  labelling axes, space group Acam was used so the c and a axes are swapped.
\end{thebibliography}

\end{document}